\documentclass[conference,10pt]{IEEEtran}
\IEEEoverridecommandlockouts

\usepackage{graphicx}
\usepackage{epstopdf}
\usepackage{amsmath}
\usepackage{amssymb}
\usepackage{amsthm }
\usepackage{epsfig}
\usepackage{times}
\usepackage{setspace}
\usepackage{bm}
\usepackage{siunitx}
\usepackage{threeparttable}
\usepackage{balance}
\usepackage{tabularx}
\usepackage{subcaption}
\usepackage{xcolor}
\usepackage{fancyhdr}

\usepackage{array}

\usepackage[T1]{fontenc}
\usepackage{times}
\usepackage[mathscr]{eucal}
\usepackage{wrapfig}
\usepackage{multirow}
\usepackage{tablefootnote}

\newcommand{\equals}{\!=\!}

\newcommand{\minus}{\!-\!}
\newcommand{\gteq}{\!\ge\!}
\newcommand{\gthan}{\!>\!}

\newcommand{\plus}{\!+\!}

\newcommand{\rel}{\mathrm{r}}
\newcommand{\e}{\mathrm{e}}

\newcommand{\dx}{\:\mathrm{d}}

\newcommand{\Ex}{\mathop{\mathbb{E}}}

\begin{document}


\title{Coded Relaying in LoRa Sensor Networks}

\author{Siddhartha S. Borkotoky, Pavan Datta Abbineni, Vatsalya Chaubey, and Sonu Rathi 

 \thanks{
This work was supported by the Science
and Engineering Research Board (SERB), Government of India, through its Start-up Research Grant (SRG) under SRG/2020/001491. The authors are with the Indian Institute of Technology Bhubaneswar, India. (E-mail: borkotoky@iitbbs.ac.in)
}

}

\maketitle
\thispagestyle{fancy}
\begin{abstract}
To enhance the reliability of  LoRa sensor networks, we propose two opportunistic forwarding protocols in which relays overhear sensor-to-gateway messages and periodically forward them to a gateway. To reduce the number of  transmissions, the relays forward XOR sums of the overheard messages. One protocol employs a single relay node, while two relays cooperate in the other. With a single relay, our proposed protocol requires fewer relay transmissions to achieve similar loss rates as conventional uncoded forwarding in which every received message is immediately forwarded. We demonstrate up to 55\% reduction in relay transmission time relative to uncoded forwarding.  The cooperative protocol provides better loss and delay  performance than the single-relay protocol, and we demonstrate up to 16\% reduction in the message-loss rate along with 33\% reduction in relay transmission times compared to conventional forwarding. 
\end{abstract}



\section{Introduction}
\label{sec:intro}
The LoRa technology has established itself as a major enabler of IoT connectivity due to its low-power operations, robustness against propagation loss, and simplicity of implementation. 
However, interference-induced frame losses  are a major concern in LoRa~\cite{GeR17}. Use of acknowledgements (ACKs) and retransmissions to combat frame loss are often infeasible since scanning for ACKs after every transmission drains sensor battery and reduces network lifetime; duty-cycle regulations prevent the gateway from sending frequent ACKs; and ACKs and retransmissions may themselves cause frame loss~\cite{PRK17}. Alternative reliability mechanisms are therefore of interest. In this paper, we propose and evaluate one such mechanism in which relay nodes are used to provide additional routes between the sensors and gateway. Relays opportunistically overhear the sensor messages and  forward them to the gateway, so that a message that did not reach the gateway directly may reach via a relay. One of our primary motives is to examine the efficacy of \textit{coded forwarding}, in which a relay forwards bitwise XORs of sensor messages, as opposed to \textit{uncoded forwarding}, in which each message is forwarded  as it is. 

Relaying in LoRa networks and its potential benefits are reported in~\cite{BSB19}--\cite{TBK20}.  Uncoded forwarding is employed in~\mbox{\cite{BSB19}--\cite{TJY20}}. In the coded relaying scheme of~\cite{TBK20}, multiple relays  employ random linear network coding to  combine received sensor messages before sending. Relays use feedback from the gateway to decide when to code and transmit, and the coding is performed in Galois Field of order 16 (GF(16)).       

By contrast, we assume unsophisticated, low-cost relays that receive periodically (as in~\cite{BSB19}) and then transmit XOR sums of received messages. Such GF(2) operations lead to low computational requirements. We assume that the gateway does not send any feedback  either to the sensors or to the relays. The sensors transmit each message only once and do not listen to downlink frames (except infrequent control messages carrying configuration data) so as to prevent battery drain. No routing protocol is required; the sensors transmit to the gateway without knowledge of the relay's presence. 


Through simulations and analysis, we demonstrate that the duration for which a relay stays in the receive mode is a crucial design parameter. With its appropriate selection, our proposed methods outperform uncoded forwarding in terms of both loss rate and the average number of transmissions by the relay.


\begin{figure} 
    \centering
    \includegraphics[scale=0.2, bb=-50 240 820 520]{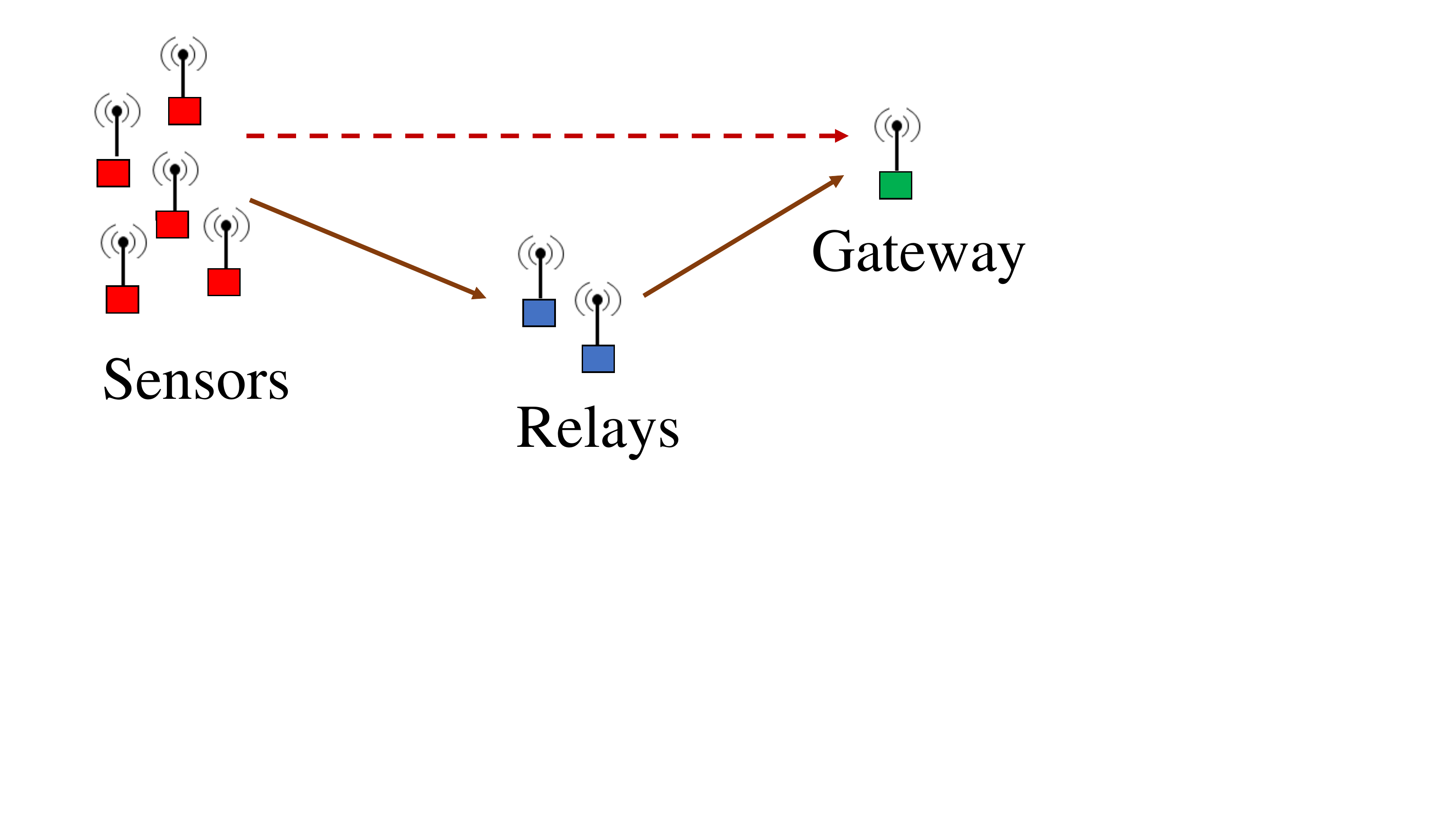}
    \setlength{\belowcaptionskip}{-12pt}
    \caption{A LoRa sensor network with relays.}
    \label{NetDiag}
\end{figure}

\section{Relaying Protocols} 
\label{sec:protocols}
We consider up to two relays placed in between a cluster of sensors and the gateway (see Fig.~\ref{NetDiag}).  
We assume slotted ALOHA transmissions by the sensors, whereas the relays transmit over predefined time slots. Time slotting is also assumed in~\cite{LZK17}--\cite{TJY20}. No frame-level synchronization is required at the sensors, which only need  to be aware of the slot boundaries. Regardless, our protocols can be easily adapted to pure ALOHA transmissions from sensors.  We refer to the two protocols as the \textit{single-relay protocol} and the \textit{cooperative protocol}, depending on whether they use one relay or two relays. In each, the sensors transmit messages to the gateway, and the relays try to overhear the sensors' frames.

\subsection{The Single-Relay Protocol}
\begin{figure} 
    \centering
    \includegraphics[scale=0.17, bb=70 0 820 520]{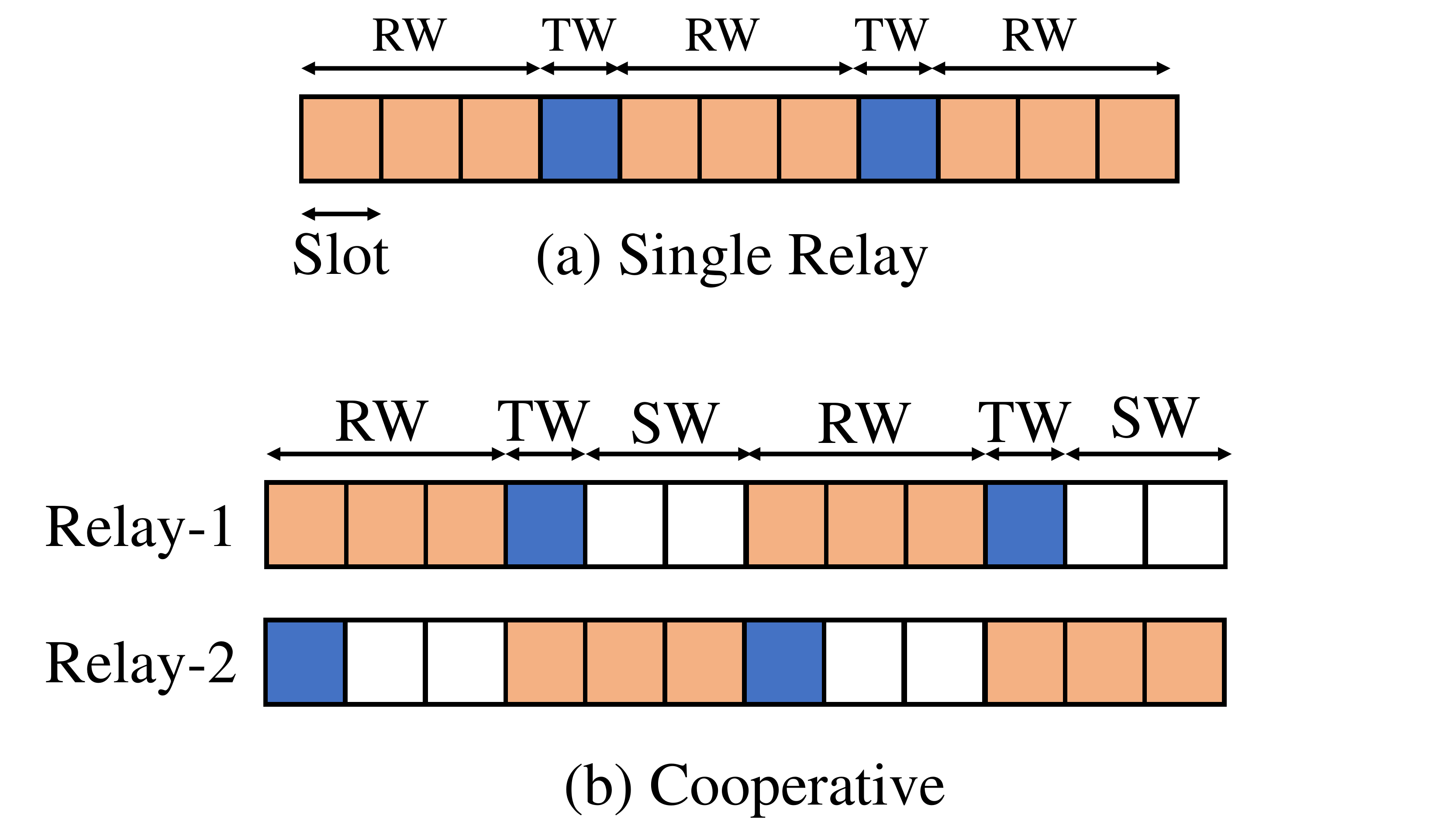}
    \setlength{\belowcaptionskip}{-12pt}
    \caption{Relay timing diagram (shown for $n_r\!=\!3$ and $n_s\!=\!2$).}
    \label{TimingDiag}
\end{figure}

In the single-relay protocol, the relay alternates between two states -- a \textit{receive window} (RW) having $n_r$ slots and a \textit{transmit window} (TW) having a single slot. This is illustrated in Fig.~\ref{TimingDiag}. In the receive window, the relay tries to overhear the sensors' transmissions and stores the received messages. In the transmit window, the relay transmits a \textit{coded frame} containing the sum (bitewise XOR) of the stored messages. The gateway can recover one message from a coded frame if all other XORed messages have already been received. Coded frames that fail to deliver a message are discarded.

\subsection{The Cooperative Protocol}
The cooperative protocol employs two relays, each having an  $n_r$-slot receive window and a \mbox{single-slot} transmit window. In addition, the relays also have a \textit{sleep window} of $n_s$ slots. As in the single-relay protocol, each relay overhears sensor frames in the receive window, and transmits  the sum of the overheard messages in the following transmit window. In the sleep window, a relay stays idle to conserve energy. The  relay windows, shown in Fig.~\ref{TimingDiag}, are such that $n_r \equals n_s \plus 1$, and one relay's receive window coincides with the other's transmit and sleep windows. This ensures that at any given instant, there is one relay available to receive.


\pagestyle{empty}
\section{System Model}
\label{sec:sys_model}
We denote the number of sensors by $n$ and the slot length by $l_s$. A sensor frame contains $b_{\mathrm{pl}}$ bytes of measurement data and a header with a $b_{\mathrm{id}}$-byte sensor ID and a $b_{\mathrm{seq}}$-byte sequence number. Since a coded message is the sum of multiple sensor messages, each  $b_{\mathrm{pl}}$ bytes long, the coded messages are also of length  $b_{\mathrm{pl}}$ bytes. In addition to the sum, the coded frame also contains the identifiers (originating node IDs and sequence numbers) of the summed messages as its payload.  

We assume the interval between two consecutive transmissions from a sensor to be exponentially distributed with parameter $\lambda$. Thus $\lambda^{-1}$ is the mean interval between two consecutive frames, which is assumed to be much larger than the slot length $l_s$. Each transmission begins at the slot boundary, and a slot is long enough to accommodate one full frame. Such assumptions are consistent with typical LoRa applications in which EDs send small amounts of data infrequently. 

The distances from the sensors to the gateway are independent and identically distributed (i.i.d.) with probability density function (pdf) $f_D(x)$. Similarly, the distances from the EDs to the relays are i.i.d. with pdf $f_{D'}(x)$. The power-fading coefficient $A$ on a link remains constant over the reception of a frame but varies from frame to frame with pdf $f_A(x)$. The cumulative distribution functions (cdfs) of $D$, $D'$, and $A$ are denoted by $F_D(x)$, $F_{D'}(x)$, and  $F_A(x)$. 

All frames are transmitted with the same power. The received power at the gateway from an ED at distance $D$  is $R \equals \gamma  A D^{-\alpha}$, where the constant $\gamma$ depends on antenna characteristics and transmit power. The cdf of $R$ is
\begin{align} 
    F_R(x) &= P(\gamma  A D^{-\alpha} < x) = \int_u F_A(\gamma^{-1} u^\alpha x) f_D(u) \dx{u}.
\end{align}
Similarly, the received power $R'$ at the relay has the cdf
\begin{align} 
    F_{R'}(x) &= \int_u F_A(\gamma^{-1} u^\alpha x) f_{D'}(u) \dx{u}.
\end{align}
For simplicity, the distance between the gateway and the relay(s) is assumed to be fixed at $d_r$. Consequently, the received power at the gateway from the relays has the cdf 
\begin{align} 
    F_{R''}(x) &= F_A(\gamma^{-1} d_r^\alpha x).
\end{align}
We denote the receiver sensitivity by $\zeta$. An incoming frame goes undetected if its power at the receiver is below $\zeta$.  

Frame duration and receiver sensitivity both depend on the \textit{spreading factor} (SF), which is an integer between 7 and 12. Larger SFs result in longer frames and lower sensitivity. The SFs are quasi-orthogonal, therefore the interference experienced by a frame from signals of different SFs is rather small. We denote by $L(s,b)$ the duration of a frame with $b$ bytes payload and SF $s$. The frame length computation is described in~\cite{Sem13}. We assume sensors and relays to use different SFs and neglect interference between relay and sensor transmissions. 

The LoRa modulation exhibits the \textit{capture effect}. In order for a frame to be correctly demodulated in the presence of other same-SF frames, its power must exceed the strongest interferer's power by at least an amount equal to the \textit{capture threshold}, which we denote by $\xi$ (in ~dB)~\cite{GeR17}.  

\section{Performance Analysis}
Our performance metrics are the \textit{message loss rate} (MLR) and the \textit{relay duty cycle} (RDC). The MLR is the probability that a sensor message fails to reach the gateway, that is, neither the \mbox{sensor-to-gateway} direct link nor a  relay is able to deliver the message. The RDC is the  fraction of time the relays spend transmitting frames. With all relay frames having the same transmit power, the RDC metric captures the energy spent by the relays in forwarding sensor messages. Since LoRa networks operate in unlicensed frequency bands, the RDC may also be interpreted as a measure of the interference contributed by the relays to other devices occutpying the same band.  


\subsection{MLR for the Single-Relay Protocol}
The MLR for the single-relay protocol is 
\begin{equation}
    \mathrm{MLR}_{\text{single-relay}} = 1 - S_{\mathrm{dir}} - S_{\mathrm{rel}},
\end{equation}
where $S_{\mathrm{dir}}$ is probability that the sensor delivers the message directly to the gateway (the \textit{direct delivery probability}), and  $S_{\mathrm{rel}}$ is the probability that the direct transmission fails but the relay is able to deliver the message to the gateway (the \textit{relay delivery probability}). We can express $S_{\mathrm{rel}}$ as
\begin{equation}
    S_{\mathrm{rel}} = S_{\text{s-}\rel}S_{\rel\text{-g}}S_{\mathrm{c}}
\end{equation}
where $S_{\text{s-}\rel}$ is the probability that the sensor's transmission fails at the gateway but is  received by the relay, $S_{\rel\text{-g}}$ is the probability that the relay's coded frame is delivered to the gateway, and $S_{\mathrm{c}}$ is the probability that of all XORed messages in the frame,  only one message is missing at the gateway. 

Consider a message from an arbitrary sensor located at distance $D$ from the gateway. We call this message the \textit{desired message}.  We will derive $S_{\mathrm{dir}}$ and $S_{\mathrm{rel}}$ for the desired message in the following subsections.

\subsubsection{Direct Delivery Probability}
Suppose that $K$ other nodes also transmitted their messages in the same slot as the desired message. Denote the received power at the gateway in the frame carrying the desired message by $R_0 \equals \gamma A D^{-\alpha}$. Conditioned on $D \equals d_0$, $A \equals a_0$, and $K\equals k$, we write the conditional probability that this frame is received over the direct link as
$S_{\mathrm{dir}}(d_0,a_0,k)$. It is
the probability that the frame survives both fading and interference, and can be written as
\begin{equation} \label{eq:Sd_cond_all3}
    S_{\mathrm{dir}}(d_0, a_0, k) = \theta(d_0,a_0)S_I(a_0,d_0,k),
\end{equation}
where $\theta$ specifies whether the desired frame survives fading. It is defined as $\theta(d_0,a_0) \equals 1$ if $\gamma a_0 d_0^{-\alpha} \gthan \zeta$, and \mbox{$\theta(d_0,a_0) \equals 0$} otherwise, where $\zeta$ is the receiver sensitivity. The term $S_I(a_0,d_0,k)$ is the conditional probability that the frame survives interference, and is equal to the probability that the received power $r_0 \equals \gamma  a_0 d_0^{-\alpha}$ exceeds the maximum of the powers in the $k$ interferers by at least $\xi$ dB. Let $\hat{R}$ be the strongest interferer power. Then
\begin{align} \label{eq:SI_cond_all3}
    S_I(a_0,d_0,k) &= P(r_0/\hat{R} > \tilde\xi)  
    = \left(F_{{R}}(\tilde\xi^{-1}\gamma a_0  d_0^{-\alpha})\right)^k,
\end{align}
where $\tilde\xi \equals 10^{0.1\xi}$.   
To find the probability that there are $k$ interferers, note that the probability that a sensor makes a transmission in any given slot is approximately 
\begin{equation}
p_{\mathrm{tx}} \approx 1 - \e^{-\lambda l_s}. 
\end{equation}
To see this, note that the exponential($\lambda$) distribution of the inter-arrival times result in a Poisson arrival process with rate $\lambda$ for each sensor. The probability that a sensor produces no measurement within a slot is $\e^{-\lambda l_s}$. Since the probability that a sensor generates multiple measurements within  a slot is negligible, $1 \minus \e^{-\lambda l_s}$ is the approximate probability that it transmits in a given slot. Thus the number of interferers $K$ is a binomial $(n \minus 1, p_{\mathrm{tx}})$ random variable, to which we employ the Poisson approximation 
\begin{equation}
    P_K(k) \approx \frac{\nu ^k \e^{-\nu}}{k!}, \quad 0 \leq k \leq n \minus 1,
\end{equation}
where $\nu \equals (n \minus 1) p_{\mathrm{tx}}$. Now using~\eqref{eq:SI_cond_all3} in~\eqref{eq:Sd_cond_all3} and deconditioning on $K$, we obtain 
\begin{align}
S_{\mathrm{dir}}(a_0, d_0) \nonumber
&= \sum_{k = 0}^{n-1}  \frac{\nu ^k \e^{-\nu}}{k!}  \theta(a_0,d_0) \left(F_{{R}}(\tilde\xi^{-1}\gamma a_0  d_0^{-\alpha})\right)^{k} \\ 
&= \theta(a_0,d_0) \e^{-\nu}\sum_{k = 0}^{n-1}  \frac{ \left(\nu\lambda_R(a_0,d_0)\right)^{k}}{k!}, \end{align}
where
\begin{align}
    \lambda_R(a_0,d_0) =  F_{{R}}(\tilde\xi^{-1}\gamma a_0  d_0^{-\alpha}).
\end{align}
Approximating the summation by the exponential function,
\begin{align} \nonumber
S_{\mathrm{dir}}(a_0,d_0) &\approx \theta(a_0,d_0) \e^{-\nu} \e^{\nu  \lambda_R(a_0,d_0)} \\ 
&= \theta(a_0,d_0) \e^{-\nu(1- \lambda_R(a_0,d_0))}.
\end{align}
The final expression for $S_{\mathrm{dir}}$ is given by 
\begin{align}
S_{\mathrm{dir}} = \Ex_{(A,D)} \left[\theta(A,D) \e^{-\nu(1- \lambda_R(A,D))}\right], 
\end{align}
where
\begin{align}
\Ex_{(A,D)}[(g(A,D))] = \int_{u} \int_{a} g(a,u) f_A(a)f_D(u)\dx{a}\dx{u},   
\end{align}


\subsubsection{Probability of Relay Delivery}
A necessary -- but not sufficient -- condition for the  delivery of a frame to the relay is that the frame must be transmitted within the relay's receive window. This occurs with probability
\begin{equation} \label{p_rw}
    p_{\mathrm{rw}} = \frac{n_r}{n_r + 1}.
\end{equation}

Now consider a frame transmitted during the relay's receive window. Let $S_{\mathrm{rx}}$ be the probability that it is lost at the gateway but is received by the relay. To find $S_{\mathrm{rx}}$, we condition on $K \equals k$ interferers in the slot, $D \equals d_0$ and $D' \equals d_1$ being the sensor-to-gateway and sensor-to-relay separations, respectively, and \mbox{$A \equals a_0$} and $A' \equals a_1$ being the fading coefficients on the sensor-to-gateway and sensor-to-relay links, respectively. We then have the following conditional reception probability 
\begin{align} 
    S_{\mathrm{rx}}^{\mathrm{cond}} &= (1-S_{\mathrm{dir}}(a_0,d_0,k))S_r(a_1,d_1,k),
\end{align}
where $S_r$ is the probability of correct reception over the sensor-to-relay link during a slot in which the relay is receiving. Its derivation is identical to that of $S_{\mathrm{dir}}$, except that the distance pdf is $f_{D'}(\cdot)$ instead of $f_{D}(\cdot)$. Therefore,
\begin{align} \label{eq:S_rx_1}
    S_{\mathrm{rx}}^{\mathrm{cond}} &= \left[1-\theta(a_0,d_0)(\lambda_{R}(a_0,d_0))^k\right]\theta(a_1,d_1)(\lambda_{R'}(a_1,d_1))^k.
\end{align}
Removing the conditioning, we obtain
\begin{align}
    S_{\text{s-}\rel} &= p_{\mathrm{rw}}  \sum_{k = 0}^{n-1}  \frac{\nu ^k \e^{-\nu}}{k!}   
     \Ex_{(A,A',D,D')} [S_{\mathrm{rx}}^{\mathrm{cond}}],
\end{align}
where,
\begin{align} \nonumber
    \Ex_{(A,A',D,D')}&[g(A,A',D,D']  \\ \nonumber
    &= \int_{a} \int_{a'}  \int_{u} \int_{v} g(a,a',u,v) \\ \nonumber & f_A(a)f_A(a')f_D(u)f_{D'}(v). \dx{a}\dx{a'}\dx{u}\dx{v}.
\end{align}
The expression above simplifies greatly if the sensor locations are such that the loss due to fading is small, and the sensors are clustered such that each of them has approximately the same distance $d_0$ from the gateway and $d_1$ from the relay. In such cases, we may substitute $\theta(a_0,d_0) \approx \theta(a_1,d_1) \approx 1$ in~\eqref{eq:S_rx_1} and remove the conditioning to obtain
\begin{align} \nonumber
    S_{\text{s-}\rel} &= p_{\mathrm{rw}} \cdot \Bigg( \Ex_{A_1} \left[\e^{-\nu(1- \lambda_R(A_1,d_1))}\right] \\
    &+ \Ex_{(A_0,A_1)} \left[\e^{-\nu(1- \lambda_R(A_0,d_0)\lambda_R(A_1,d_1))}\right]\Bigg),
\end{align}
where $A_0$ and $A_1$ both have the pdf $f_A(x)$. 

To find $S_{\rel\text{-g}}$, the success probability on the relay-to-gateway link, only frame loss due to fading is of concern since the relay uses a different spreading factor than the sensors. Hence,
\begin{equation}
    S_{\rel\text{-g}} = 1 - F_A(\gamma^{-1} \zeta \tilde{d}_r^{\alpha}),
\end{equation}
where $\tilde{d}$ is the relay-to-gateway separation. 

The only parameter left to find is $S_{c}$, the probability that the gateway is able to recover the desired message from a coded message. Since recovery is possible if and only if all XORed messages other than the desired message have already been delivered to the gateway, we wish to derive the probability that in the preceding receive window, the relay received $m$ messages (other than the desired message), each of which was also received by the gateway. Using the usual conditioning and deconditioning, the probability that a frame transmitted within the relay's receive window is received by both the relay and the gateway is
\begin{align}
    p_{\mathrm{gr}} \equals \sum_{k=0}^{n-1}\frac{\nu^k \e^{-\nu}}{k!}\Ex_{(A_0,A_1,D,D')}[S_{\mathrm{dir}}(A_0,D,k)S_r(A_1,D',k)].
\end{align}
We can now determine $S_c$ as follows
\begin{align}
    S_c = \sum_{m=0}^{n_r - 1} {n_r - 1 \choose m} p_{\mathrm{gr}}^{m} f^{n_r - m - 1},
\end{align}
where $f$ is the probability that no message is successfully received at the relay over a given slot. It is given by 
\begin{align} \label{eq:f} 
f = 1 - n p_{\mathrm{tx}} \Ex_{(A,D')} \left[\theta(A,D') \e^{-\nu(1- \lambda_R(A,D'))}\right]. 
\end{align}
The product of $p_{\mathrm{tx}}$ and the expectation in~\eqref{eq:f} is the probability that an arbitrary sensor transmits a message that is correctly received by the relay. The multiplier $n$ accounts for the fact that there are $n$ such sensors.  

\subsection{RDC of the Single-Relay Protocol}
Suppose that the relay received $m \gteq 1$ messages in the receive window preceding a certain transmit window. Then it must send a coded frame of length $b_{\mathrm{pl}}$ bytes along with a header containing sender ids and sequence numbers of the messages that are XORed. Assuming all these to be part of the relay payload, the payload length in bytes is 
\begin{equation}
  b_c(m) = b_{\mathrm{pl}} + m(b_{\mathrm{id}} + b_{\mathrm{seq}}).  
\end{equation}
The average relay frame length is therefore 
\begin{equation}
    l_{\mathrm{av}} = \sum_{m=1}^{n_r} {n_r \choose m} (1-f)^m f^{n_r - m} L(\mathrm{sf}_r, b_c(m)), 
\end{equation}
where $\mathrm{sf}_r$ is the SF used by the relay. The average transmission duration per transmit window is hence $l_{\mathrm{av}}$, and the fraction of time the relay spends transmitting is \begin{align}
    \mathrm{RDC}_{\text{single-relay}} = \frac{l_{\mathrm{av}}}{(n_r + 1)l_s}.
\end{align}

\subsection{Analysis of the Cooperative Protocol}
The MLR analysis for the cooperative protocol is the same as that of single-relay case if we assume that the distance from the gateway to both relays are approximately the same, and so are the pdfs of the distances from the sensors to the two relays. We may then visualize the two relays as a single full-duplex relay that is always receiving. The MLR for can then be analyzed using an approach identical to the one for the single-relay protocol, except that we must use ${p_\mathrm{rw}} \equals 1$.

For the RDC too, the derivation is identical except that now there are two relays and each sleeps for $n_s$ slots in addition to transmitting and receiving. The resulting expression is 
\begin{align}
    \mathrm{RDC}_{\mathrm{cooperative}} = \frac{2l_{\mathrm{av}}}{(n_r + n_s + 1)l_s}.
\end{align}
Since $n_r \equals n_s \plus 1$, the above simplifies to
\begin{align}
    \mathrm{RDC}_{\mathrm{cooperative}} = \frac{ l_{\mathrm{av}}}{(n_r + 0.5)l_s}.
\end{align}

\section{Performance Results}
We first consider a network of 20 sensors, each having a 1-byte ID, and transmitting 10-byte messages with a mean inter-frame arrival time of 17.5~s. Each frame has a 1-byte sequence number.  The sensors and relays use SF 8 and 7, respectively. The links have block Rayleigh fading. Our performance graphs show the analytical results, which we found to be in very close agreement with simulations.  

\begin{figure} 
    \centering
    \includegraphics[scale=0.26]{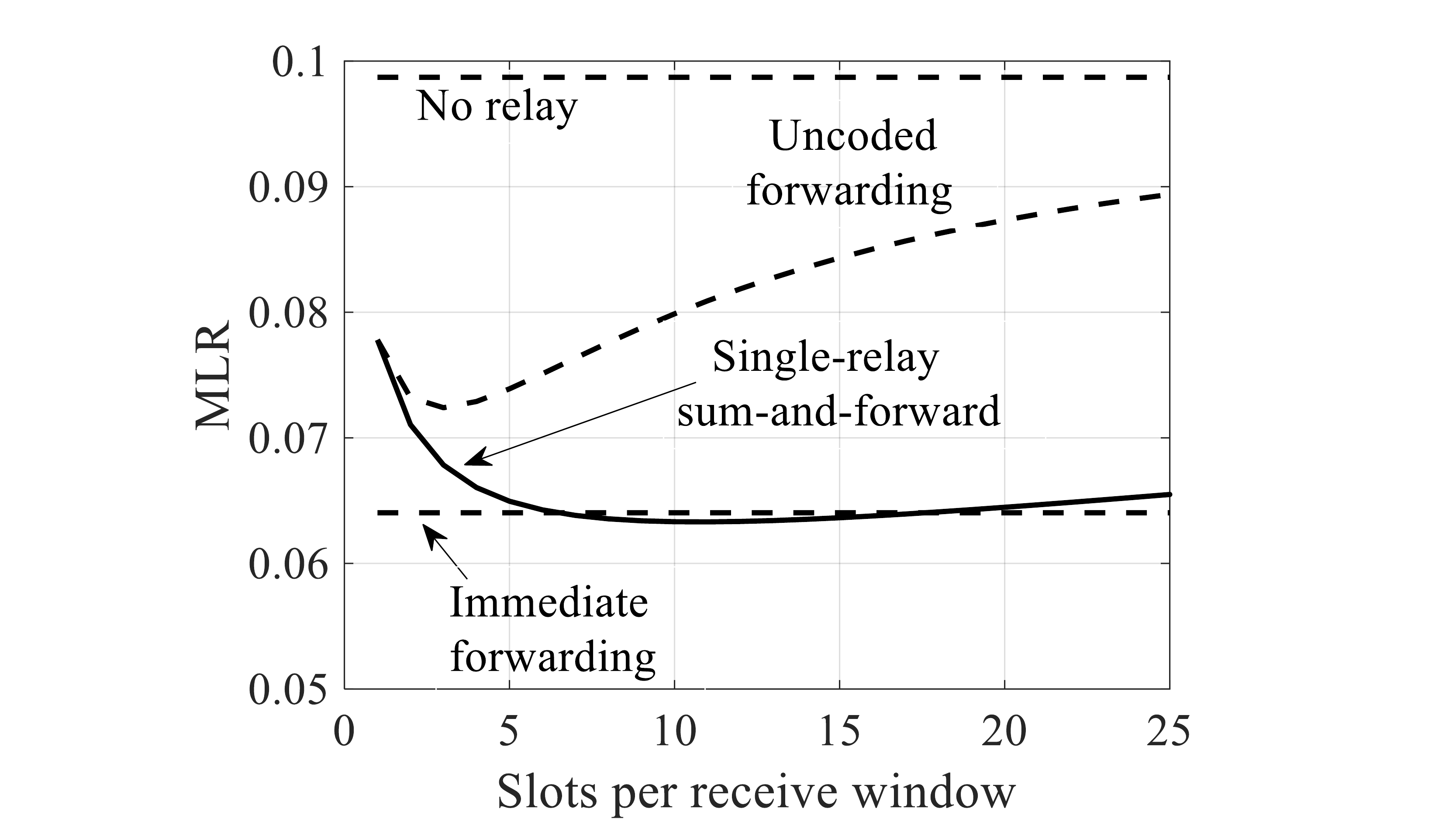}
   \setlength{\belowcaptionskip}{-12pt}
    \caption{MLR comparison of various protocols.}
    \label{Fig3_new}
\end{figure}

Fig.~\ref{Fig3_new} compares the MLR for the single-relay protocol with three baselines: (1) The \textit{no-relay} protocol does not employ relays. (2) The \textit{immediate-forwarding} protocol uses a single relay that forwards every received message immediately in the following slot. This is essentially the protocol of~\cite{BrC20}. (3)  The \textit{uncoded-forwarding} protocol is same as our proposed single-relay protocol except that the relay forwards the individual messages received during the previous receive window instead of transmitting their sum. If all messages do not fit within the transmit window, some messages are discarded at random. This approach is identical to that of~\cite{BSB19} with one relay. To distinguish our proposed single-relay protocol from the benchmarks that also employ one relay, we occasionally refer to it as the \textit{single-relay sum-and-forward} protocol.     

It is clear from Fig.~\ref{Fig3_new} that any form of relaying provides better MLR than the system without a relay, even though each relay-based protocol compared here has only a single relay against 20 sensors. The no-relay and immediate-forwarding  protocols do not have the concept of $n_r$, so we plot their MLRs as straight lines. The uncoded-forwarding protocol -- although better than no-relay -- performs poorly relative to the other relay protocols, especially for larger receive window sizes. A long receive window results in more received messages than the relay can send in the transmit window slot, resulting in dropped massages. By contrast, our proposed single-relay protocol performs significantly better, despite having the same transmit window duration. This is the benefit of the sum-and-forward approach, which only needs a transmit window long enough to accommodate one XORed message (whose length is the same as any of the individual information messages) and the necessary header information. Like uncoded forwarding, the MLR of the proposed protocol is sensitive to the receive window size. Having too small a receive window often leads to the relay not being able to receive any frames during a receive window, and consequently having to sit idle in the next transmit window. Having too large a receive window means that the relay's transmitted frames are XORs of a large number of messages, and the probability that the gateway is missing only one of them is small. The optimal receive window for this scenario is found to be of 11 slots. 

\begin{figure} 
    \centering
    \includegraphics[scale=0.26]{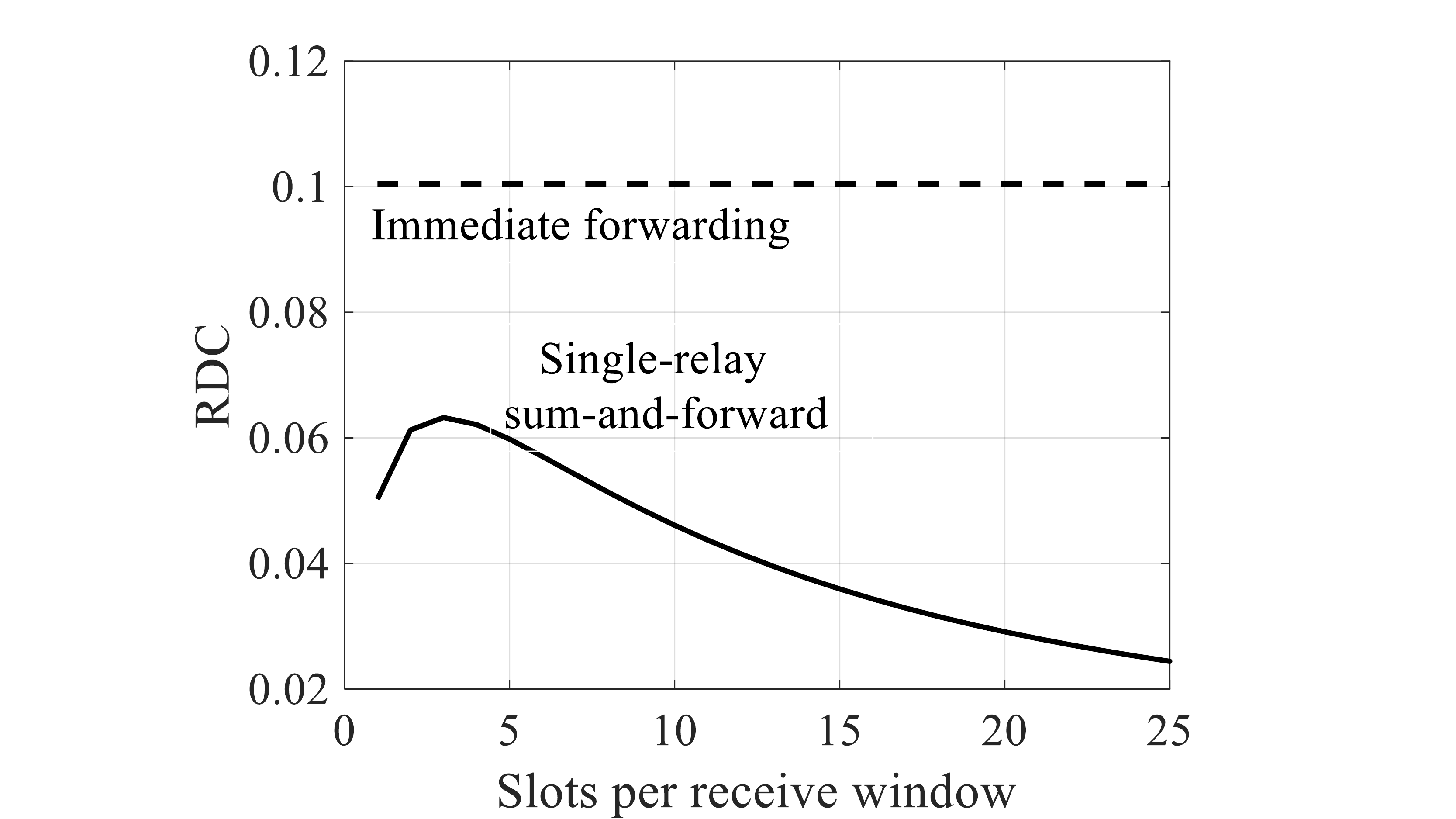}
   \setlength{\belowcaptionskip}{-12pt}
    \caption{RDC comparison of the proposed single-relay protocol and immediate forwarding.}
    \label{Fig4_new}
\end{figure}

One might conclude from Fig.~\ref{Fig3_new} that the proposed sum-and-forward approach hardly provides any benefit over immediate forwarding. But Fig.~\ref{Fig4_new} demonstrates the advantage of the sum-and-forward approach by comparing the respective RDCs, which are significantly higher for immediate forwarding. For example, with $n_r \equals 11$, immediate forwarding requires 42\% more relay transmission time than our proposed technique. Thus, single-relay sum-and-forward would ensure  significantly longer operation of battery-powered relays and also contribute less interference to the network. 



\begin{figure}
     \centering
     \begin{subfigure}[b]{0.48\textwidth}
         \centering
         \includegraphics[width=\textwidth]{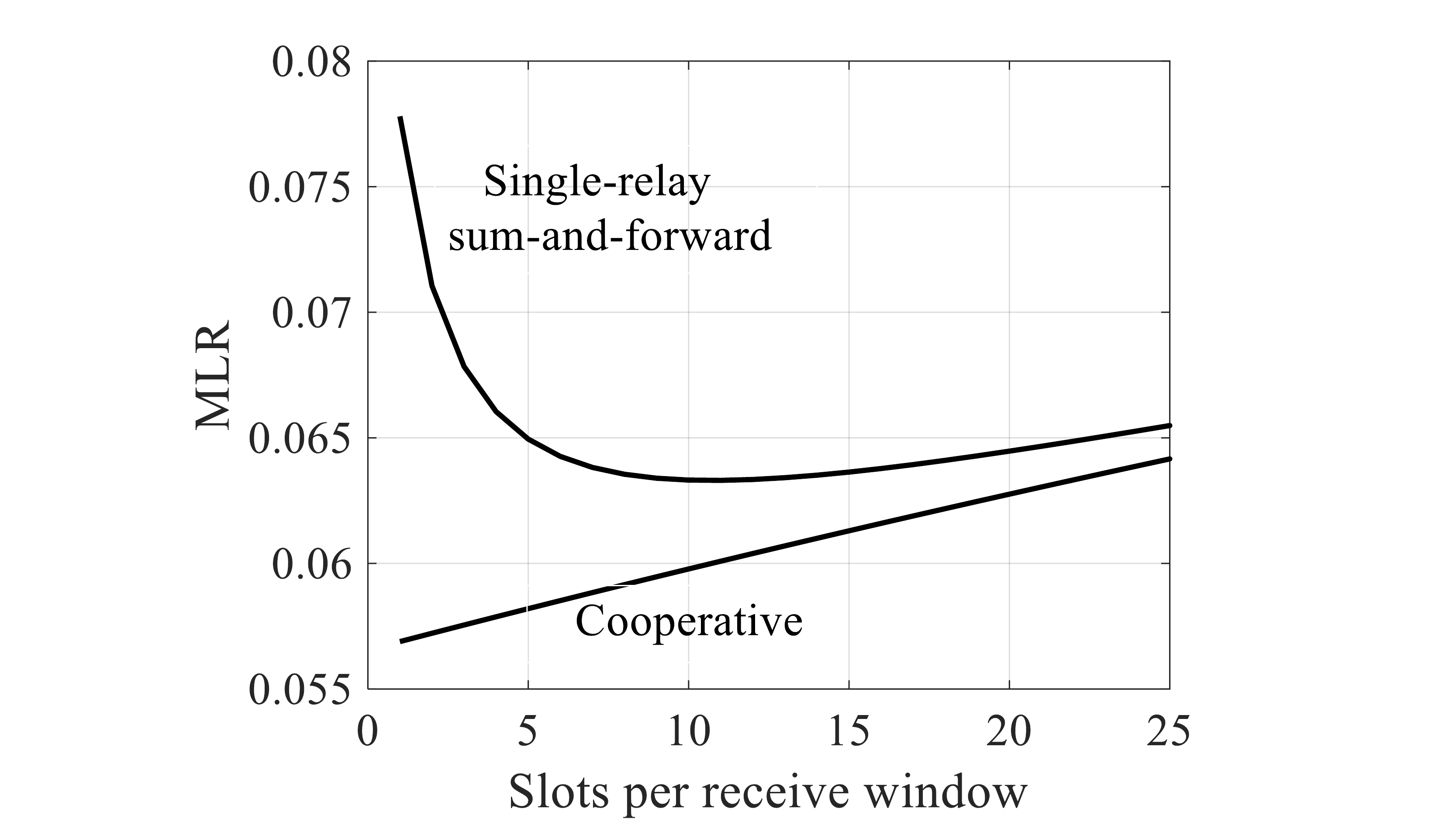}
         \caption{MLR.}
         \label{Fig5a_new}
     \end{subfigure}
     \hfill
     \begin{subfigure}[b]{0.48\textwidth}
         \centering
         \includegraphics[width=\textwidth]{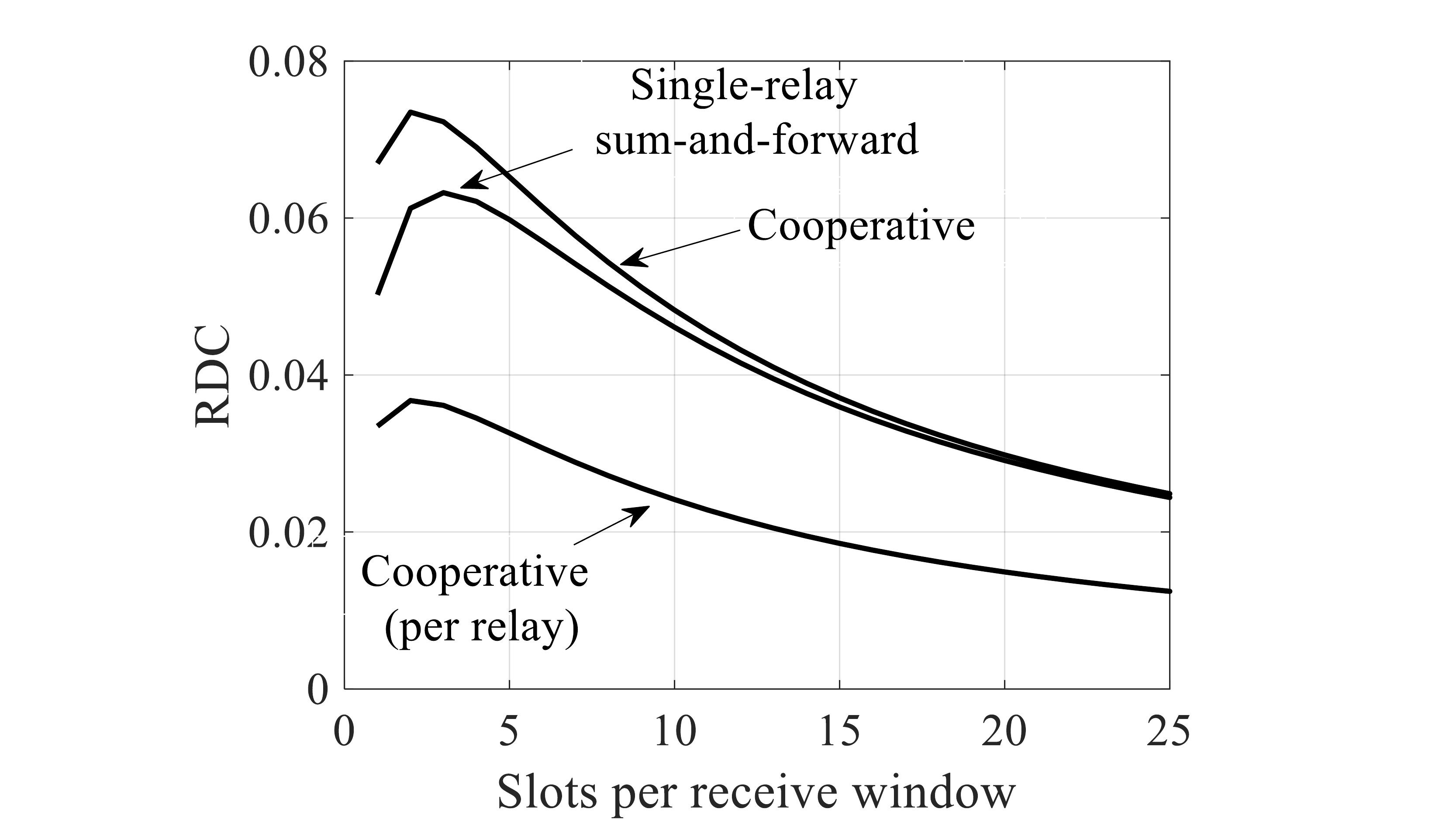}
         \caption{RDC.}
         \label{Fig5b_new}
     \end{subfigure}
     \caption{Comparison of the proposed protocols.}
     \setlength{\belowcaptionskip}{-12pt}
\end{figure}



Fig.~\ref{Fig5a_new} compares the MLR for the two proposed protocols. The cooperative protocol is able to significantly reduce the MLR relative to the single-relay approach, especially for short receive windows. For $n_r \equals 1$, the cooperative protocol provides the lowest MLR, which is 27\% lower than than that of the single relay protocol. Since $n_r \equals 1$ means that a relay has a maximum of one message to send per transmit window, there is no scope for coding in this case. By contrast, if only one relay is available, it is beneficial to forward sums of messages with an appropriately chosen receive window size.      

In addition to the MLR, the receive window size also determines the delay characteristic of the protocol, since it specifies  how long the gateway has to wait before possibly recovering a lost message via a relay. With an optimal receive window comprising a single slot, it is clear the cooperative protocol provides better delay performance than single-relay sum-and-forward at the cost of an extra relay.

Fig.~\ref{Fig5b_new} compares the RDCs of the proposed protocols. The RDC for the cooperative protocol exceeds that of the single-relay protocol for all receive window sizes. However, for the cooperative protocol, the RDC includes transmissions by both relays; the RDC per relay in this case is much smaller than the RDC for the single-relay protocol. 

\begin{figure}
     \centering
     \begin{subfigure}[b]{0.48\textwidth}
         \centering
         \includegraphics[width=\textwidth]{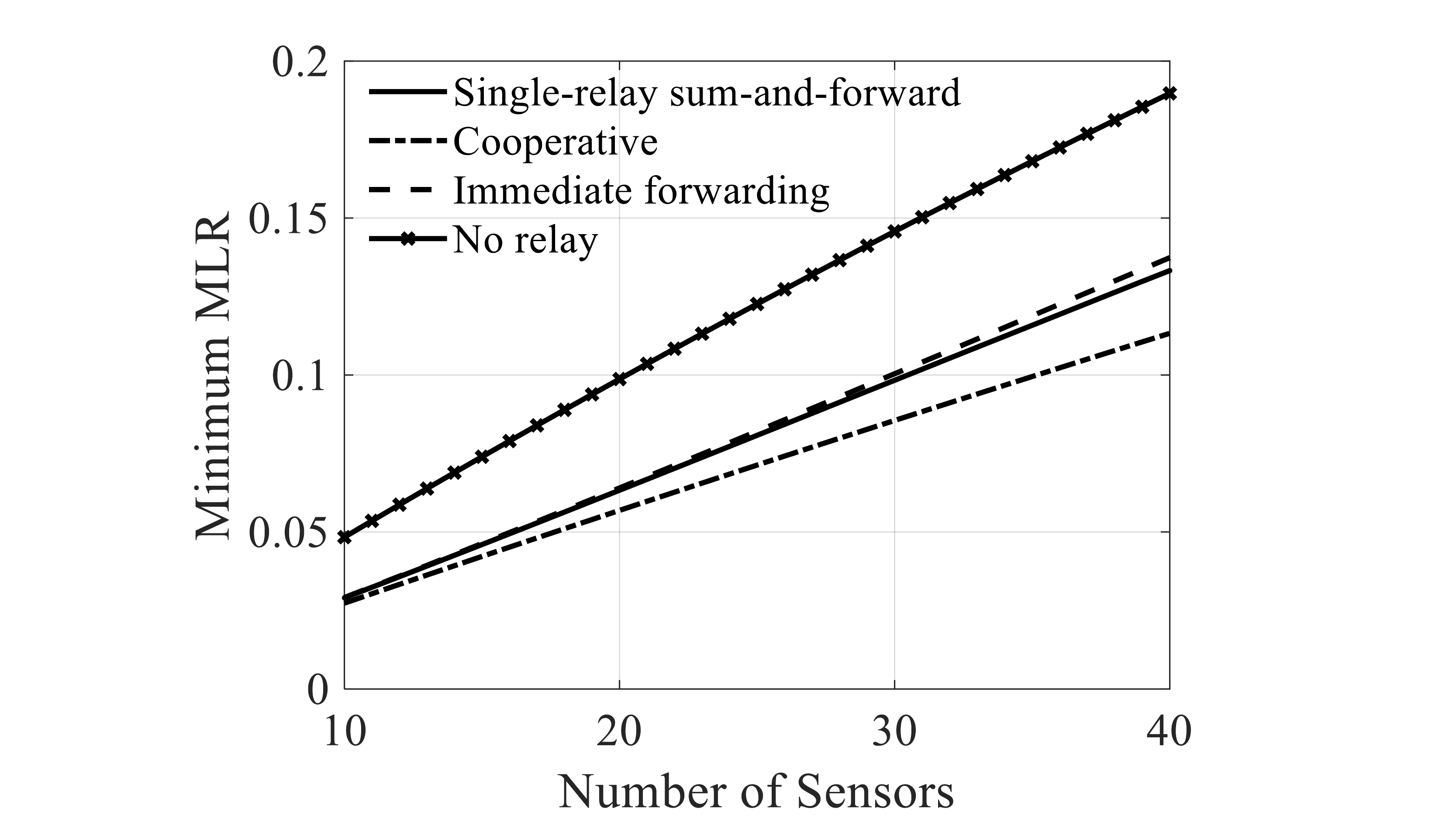}
         \caption{MLR.}
         \label{Fig6a_new}
     \end{subfigure}
     \hfill
     \begin{subfigure}[b]{0.48\textwidth}
         \centering
         \includegraphics[width=\textwidth]{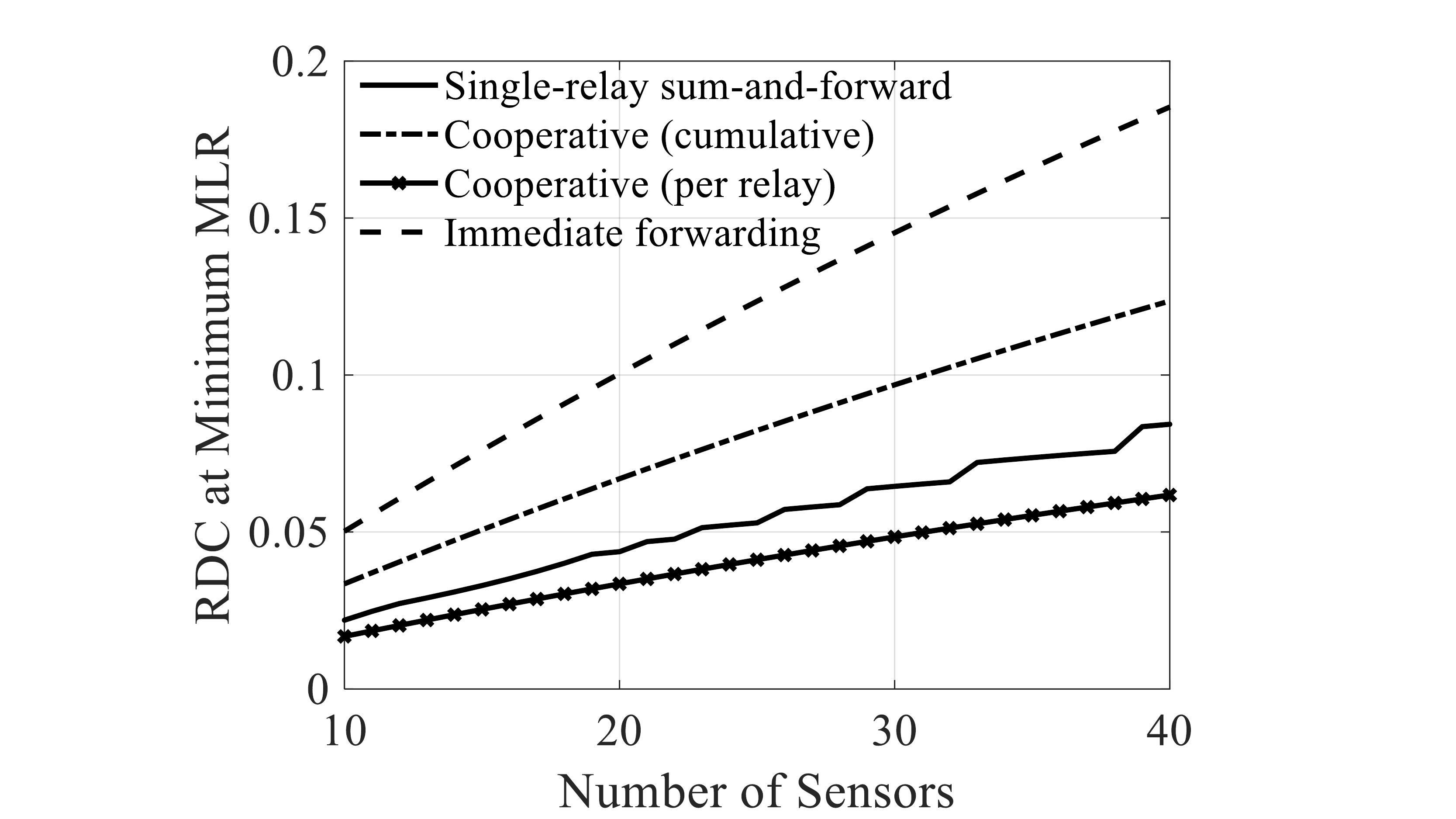}
         \caption{RDC.}
         \label{Fig6b_new}
     \end{subfigure}
     \caption{Protocol performance vs. number of sensors.}
     \setlength{\belowcaptionskip}{-20pt}
\end{figure}

Fig.~\ref{Fig6a_new} plots the MLR against the number of sensors. For the proposed protocols, the minimum achievable MLR (with respect to $n_r$) is plotted. For the no-relay and immediate-forwarding protocols, no such minimization is needed since the MLR is fixed for a given sensor count. For each protocol, the MLR increases with the number of sensors due to increased interference. The cooperative protocol provides the lowest MLR, while the immediate-forwarding and single-relay sum-and-forward protocols provide comparable results. However, as seen from Fig.~\ref{Fig6b_new}, immediate forwarding requires more transmission time at the relay (i.e., higher RDC) compared to single-relay sum-and-forward. For 40 sensors, single-relay sum-and-forward requires about 55\% less relay transmission time compared to immediate forwarding. The RDC for the cooperative protocol is also significantly lower (33\% with 40 sensors) than immediate forwarding, despite using two relays.   
 
\begin{figure} 
    \centering
    \includegraphics[scale=0.26]{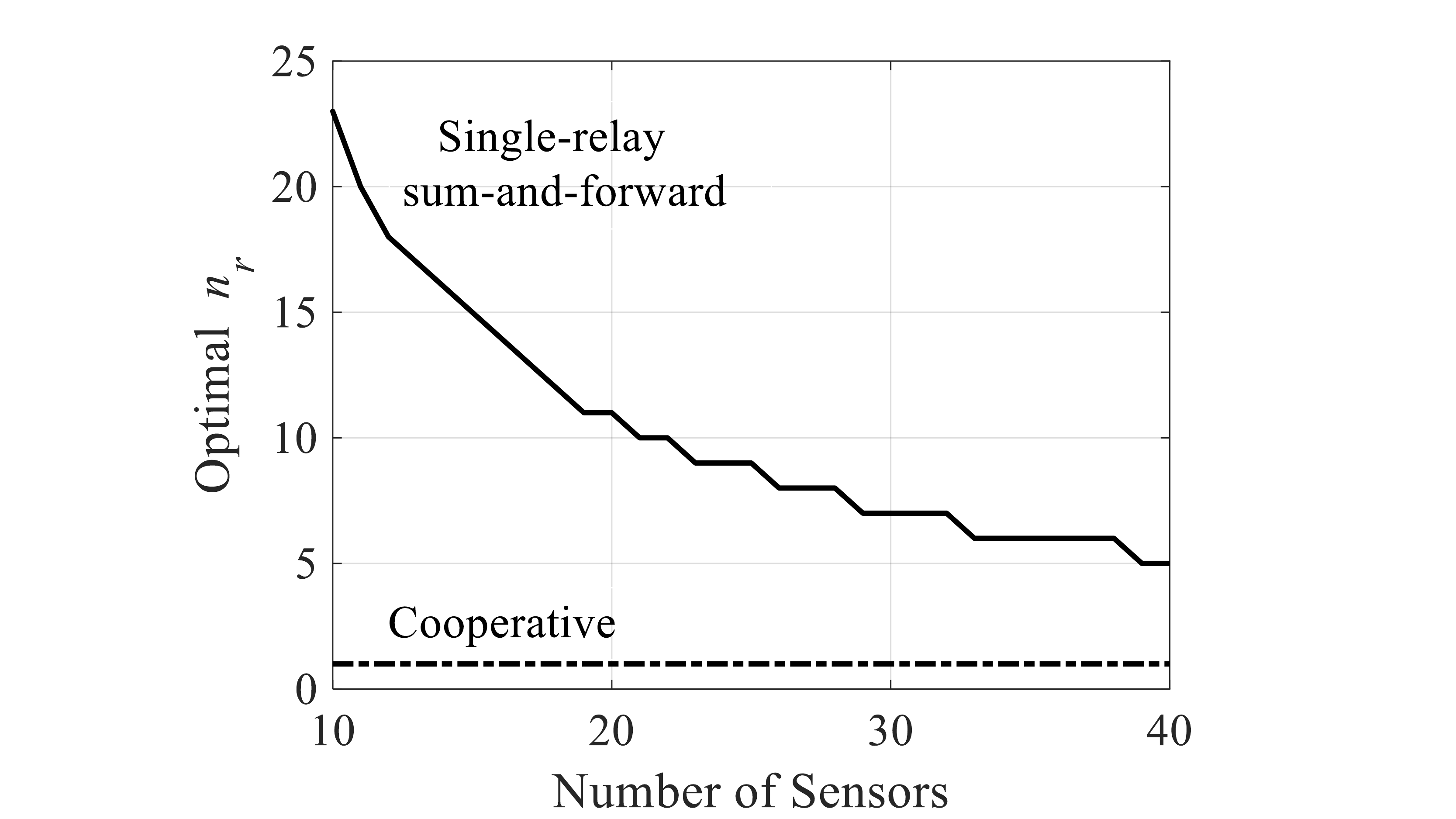}
   \setlength{\belowcaptionskip}{-12pt}
    \caption{Optimal receive window size vs. number of sensors.}
    \label{Fig7_new}
\end{figure}

The $n_r$ value that provides the lowest MLR for the proposed  protocols are shown in Fig.~\ref{Fig7_new}. We observe that $n_r \equals 1$  is the optimal choice for the cooperative protocol irrespective of the number of sensors. By contrast, the optimal receive window size for the former decreases with increasing sensor count. With large number of nodes, loss rates are high on the links. For coding to be beneficial, gateway must have exactly one missing message out of all XORed messages in a coded frame. With greater frame loss rates, coded frames containing fewer messages are more likely to be successful. A shorter receive window leads to more such coded frames.   

\section{Conclusion}
We proposed and evaluated two low-complexity relaying protocols for LoRa sensor networks. Using either one or two relays, the protocols improve the message delivery performance of the network. The proposed transmission schemes achieve similar (with a single relay) or better (with two relays) loss performance compared to conventional uncoded forwarding, while requiring fewer transmissions by the relay(s). 

\vspace{1mm}
\section{Future Work}
\vspace{1mm}
In this paper, we restricted attention to a maximum of two relays and 40 sensor nodes. As the sensor count increases, it may be necessary to employ more relays. By partitioning the sensors to multiple clusters, and assigning one or two relays to each cluster, our proposed protocols can be scaled to large networks. Devising such approaches is left as future work.

Adaptation of the protocols to pure ALOHA channel access by sensors is also of practical importance. With  pure ALOHA, the receive window of a relay may end while the reception of a sensor frame is still in progress. Analysis of relaying with pure ALOHA transmissions by the sensors is provided in~\cite{BSB19} under the assumption of periodic sensor transmissions and uncoded relaying. Analysis of coded relaying with possibly aperiodic sensor frames is left as future work.


\end{document}